\documentclass[11pt, letterpaper]{article}
\usepackage{rin} 
\usepackage[T1]{fontenc}

\title{Practical Software Approach to Digital Pulse Processing}

\author{
  Jing Liu\thanks{\url{mailto:jing.liu@usd.edu}}
}
\affil{University of South Dakota, 414 E. Clark St., Vermillion, SD 57069}

\date{Created: September 26, 2024}

\begin{document}

\maketitle
\thispagestyle{fancy} 

\begin{abstract}
  This paper presents a practical approach to digital pulse processing, emphasizing simplicity and efficiency. We advocate for a balanced software design, flat data structures, the use of the ROOT C++ interpreter, and a combination of command-line and graphical interfaces. By adopting these strategies, researchers can effectively apply digital pulse processing techniques to their specific domains without excessive theoretical concerns.
\end{abstract}

\tableofcontents

\section{Introduction}
The fundamental step in digital pulse processing involves transforming continuous analog signals into discrete digital representations. This conversion is achieved through the use of Analog-to-Digital Converters (ADCs). Prior to the invention of transistors in the 1950s, ADCs were often bulky and cumbersome. The subsequent development of transistors enabled the creation of logic gates, the foundational elements of digital circuits. A significant breakthrough occurred in the 1960s with the introduction of integrated circuits (ICs), which integrated numerous transistors onto a single silicon wafer. This innovation dramatically reduced the size and cost of ADCs.  Significant progress of ADC technology has been made in the fields of communication and imaging ever since, which also benefits the field of radiation and particle detection.

\begin{figure}[htpb]\centering\label{f:pls}
  \includegraphics[width=0.75\linewidth]{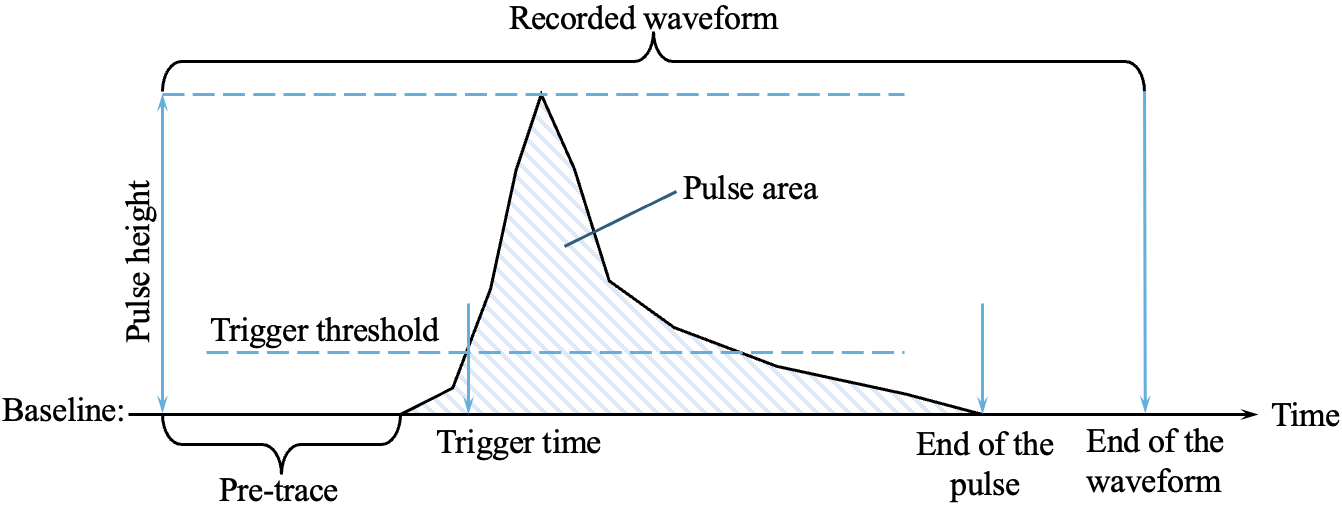}
  \caption{Basic terminology used in digital pulse processing.}
\end{figure}

A distinguishing characteristic of electronic signals generated by radiation and particle detector systems is their random occurrence in time. Consequently, these signals are referred to as pulses. To effectively capture these pulses, a triggering system is essential. This system alerts the ADC to digitize a certain length of waveform when a pulse in that waveform exceeds a pre-determined threshold as shown in Figure~\ref{f:pls}. A digitizer, a specialized electronic device, is designed to detect randomly occurring pulses through a triggering mechanism, sample waveforms that include these pulses using an ADC, and store the digitized waveforms in its onboard memory. Subsequently, these digitized waveforms can be transferred to a computer for further analysis.

To effectively extract meaningful information from the digitized waveforms, specialized software is necessary. 
First, the software needs to incorporate functions traditionally performed by analog signal processing electronics. These functions include baseline removal (adjusting the baseline of pulses to zero), pulse shaping (modifying the pulse shape for easier analysis), and denoising (reducing the impact of electronic noise on signal pulses). These software-based techniques are referred to as digital pulse processing. Second, the software needs to provide algorithms to extract key characteristics from the processed waveforms, such as pulse height, area, time above threshold, duration, and more, which are collectively referred to as digital pulse-shape analysis (PSA).

Pulse-shape discrimination (PSD) is a specialized technique within PSA that categorizes pulses based on their characteristic features. These features are indicative of the underlying interactions that produced the pulses. For instance, voltage pulses generated by alpha interactions on the surface of a high-purity germanium (HPGe) detector exhibit significantly longer rise times compared to those induced by $\gamma$-ray interactions within the bulk of the detector~\cite{alpha}; Current pulses resulting from multi-site interactions in a HPGe detector often display more fluctuations than those from single-site interactions\cite{gerda}; Pulses induced by nuclear recoils in scintillation detectors generally possess distinct characteristics compared to those induced by electronic recoils~\cite{psd}. PSD can be as straightforward as calculating the ratio of pulse height to area or fitting the decay constant of a pulse, or as complex as employing machine learning algorithms that leverage a multitude of pulse features~\cite{psd}, some of which may not be readily apparent to human observation.

Digital pulse processing offers several significant advantages over its analog counterpart. First, it provides unparalleled flexibility in selecting and implementing various processing algorithms. In contrast, analog pulse processing is often constrained by the limitations of available hardware. Second, digital processing preserves the original pulse data, allowing for the reversal of unsatisfactory processing steps. Finally, many algorithms that are difficult or impossible to implement accurately in hardware can be executed precisely in the digital domain. For these reasons, it is generally advantageous to digitize signals from radiation detectors as early as possible. The software employed to perform basic pulse processing functions becomes an indispensable tool in this context.

This article presents a compilation of lessons learned from my personal experiences in designing, constructing, and utilizing digital pulse processing software. I share insights that aim to simplify the use of such software, particularly for small-scale experiments focused on rapid research and development (R\&D). While these recommendations may not align perfectly with mainstream approaches often adopted in large-scale experiments, they can be valuable for enhancing the efficiency and effectiveness of digital pulse processing in bench-top applications up to mid-sized experiments, providing practical alternatives for those with limited resources.

To illustrate these core ideas with concrete examples, two pieces of software are introduced: TOWARD -- Tools, Objects for Waveform Analysis, Reformatting and Drawing~\cite{toward}, and ROSA -- Rootifying Output from Struck ADCs~\cite{rosa}. The former is designed to work specifically with digitizers from CAEN SpA~\cite{caen}, the latter functions for digitizers from Struck Innovative Systeme~\cite{struck}.

\section{All-in-One Versus Modularization}
\subsection{Generic Framework}
The contemporary landscape of software development is characterized by a proliferation of frameworks designed to accommodate a wide range of use cases. While this approach can lead to the creation of comprehensive software,  it is important to recognize that the resulting software is often complex and difficult to maintain, and many of these frameworks are utilized by relatively small groups of users. The substantial time and effort invested in designing generic frameworks may not always be justified, especially when considering the limited number of individuals who benefit from them.

To avoid the temptation to create a universal framework capable of interfacing with all available digitizers, I intentionally developed two separate pieces of software specifically tailored to CAEN and Struck digitizers, respectively. This approach aligns with the Unix philosophy of ``Make each program do one thing well. To do a new job, build afresh rather than complicate old programs by adding new features''.

\subsection{Modularization}
Some PSA code segment in TOWARD is duplicated in ROSA, such as pulse area integration. This is a necessary trade-off to avoid combining TOWARD and ROSA into a single framework. 

An alternative approach to avoid code duplication involves modularizing the design to isolate the integration operation into a standalone package. However, this introduces additional complexity by requiring users to install multiple packages for tasks that could potentially be accomplished with a single package. This highlights a common criticism of the Unix philosophy: excessive modularization can sometimes lead to inefficient programs.

In some cases, a moderate amount of code duplication may be acceptable if the time required to redesign the system to avoid it significantly outweighs the effort involved in copying and pasting existing code.

\subsection{Balance}
An effective software design must strike a balance between two opposing extremes: excessive consolidation and excessive modularization. Placing all functionality within a single package can result in unnecessary complexity, making the software difficult to design, maintain, and use. On the other hand, splitting the software into excessively small modules can lead to inefficiencies and complications when assembling a functional application.

Determining the optimal balance between consolidation and modularization often requires practical experimentation rather than theoretical analysis. A pragmatic approach involves starting with a focused program designed for a specific application, disregarding concerns about compatibility or code reusability. For instance, TOWARD was initially developed to handle a limited number of CAEN digitizers without considering other types. In small-scale research and development groups, this focused approach may be sufficient to meet the needs of most projects.

As new applications arise, additional programs can be created with the same targeted mindset, avoiding premature considerations of similarity or code reuse. This iterative process allows for insights into recurring patterns and shared code segments. By analyzing these patterns, it becomes possible to assess whether extracting a common component into a standalone package is justified.

This bottom-up approach, driven by real-world experience, often proves more effective than a top-down approach based solely on theoretical considerations. TOWARD and ROSA were developed following the bottom-up approach. The question of whether they represent the ideal balance between consolidation and modularization may only be answered with experience accumulated from writing similar software for more types of digitizers.

The importance of a practical, iterative approach to software design is often overlooked by those who primarily rely on theoretical principles of modularization and well-defined interfaces. While these concepts are undoubtedly valuable, real-world experience is indispensable for developing effective modularization strategies tailored to specific application domains.

Gaining insights into meaningful modularization often requires the creation and use of suboptimal software. By writing and working with less-than-ideal programs, developers can acquire valuable knowledge about the challenges and opportunities associated with different design choices. Fear of producing "bad" software should not hinder experimentation and learning. Even a seemingly flawed program can be useful for small-scale applications and serve as a foundation for building more robust and effective solutions in the future.

\section{Programming Languages}
Given the vast array of programming languages available, it's essential to select those that align with your familiarity, accessibility, and task requirements. For TOWARD and ROSA, I chose C++ for data processing, Python for the graphical user interface (GUI), and Bash for submitting batch jobs on high-performance computing (HPC) clusters. This decision was based on my proficiency in C++, the necessity of Bash for HPC job submission, and the perceived ease of developing a cross-platform GUI using Python.

\subsection{High-level Versus Low-level}
\label{s:hvl}
Programming languages can be broadly categorized as low-level and high-level. Low-level languages, while generally faster, require programmers to manage certain low-level details manually, such as compiling source code into machine-readable instructions. High-level languages, on the other hand, automate many of these tasks, resulting in increased developer productivity but potentially sacrificing some execution speed.

C++ is often considered a low-level language, chosen for its performance advantages. Python, in contrast, is a high-level language that simplifies development by eliminating the need for compilation. This characteristic makes Python more accessible to beginners compared to C++.

The boundaries between low-level and high-level languages are increasingly blurred as developers seek to combine the advantages of both. Julia, for example, is a language and system that offers automatic compilation and execution, balancing speed with user-friendliness. In Python, efficiency-critical sections of code are often implemented in C or C++ for performance optimization.  ROOT~\cite{root}, a C++-based data analysis framework, incorporates a built-in C++ interpreter called cling~\cite{cling}. This interpreter provides an interactive prompt similar to Python, allowing users to enter C++ code and receive immediate results without the need for manual compilation.

The C++ components of TOWARD and ROSA leverage the ROOT framework, benefiting from its interpreter, cling. This enables all C++ files within these projects to be executed directly by ROOT without the need for manual compilation. Following the terminology commonly used in Python and Bash, these C++ files are referred to as scripts.

For example, the following command executes the C++ function \lstinline{void idx(const char* input_file, const char* index_folder)} defined in \lstinline{idx.C} file within ROSA:
\begin{lstlisting}[language=bash]
$ root idx.C'("input.bin", "output/folder/")'
\end{lstlisting}
This command launches the ROOT interpreter and loads the \lstinline{idx.C} file, making the \lstinline{idx} function available for execution.  Strings in the single quotes are arguments passed to the \lstinline{idx} function.

\subsection{Existing Libraries}
The availability of existing libraries in a particular programming language can significantly enhance development efficiency. In addition to its interpreter, ROOT offers extensive libraries for statistical analysis of large datasets, making it well-suited for PSA.

For the graphical user interfaces of TOWARD and ROSA, I selected the tkinter~\cite{tkinter} library in Python. This library is included with standard Python installations on Windows and Mac and can be easily installed using package managers on Linux. While tkinter may have limitations compared to more comprehensive GUI libraries like PyQt, it offers a simpler and more straightforward approach for small-scale projects.

\section{Data Structure}
\label{s:struct}
A data structure designed for a limited set of use cases often offers greater simplicity and maintainability. This can reduce the effort required for creation, learning, and maintenance. While more sophisticated data structures may provide enhanced encapsulation capabilities, they can also be more complex to design, maintain, and use.  TOWARD and ROSA employ relatively simple data structures based on the following considerations.

\subsection{Multi-tiered Approach}
Data acquisition (DAQ) software accompanying digitizers typically saves output data as binary files. These files often include a header containing metadata, such as digitizer type, timestamps, trigger settings, and multiple blocks of 32-bit integers representing digitized samples. The block size is typically limited by the digitizer's internal memory. For a 10-bit ADC, analog signals are converted to integers ranging from 0 to $2^{10}=1024$. Each integer is known as an ADC count. Since a 32-bit integer can store three 10-bit ADC counts (or two for a 14-bit ADC), this encoding scheme helps conserve disk space. Binary files are frequently compressed further to reduce size. These files constitute the initial tier (tier 1) of the data processing chain.

Direct analysis of binary files from DAQ software can be cumbersome due to compression and encoding. Converting these files into a more analysis-friendly format is essential. The converted files retain much of the information from the binary files and form the second tier (tier 2)of the process chain. They are particularly useful for visualizing individual recorded waveforms.

Parameters extracted from various PSA algorithms can be saved in files without the original waveforms. Quality and physical criteria can be applied to filter out uninteresting events. These files (tier 3), significantly smaller than Tier 2 files, are suitable for statistical analysis and plot generation.

Tier 1 files can be backed up to slower storage media like tapes or online storage. Tier 2 files, while valuable for data exploration, may not require long-term retention. Tier 3 files, due to their small size, can be version-controlled alongside analysis code.

\subsection{File Format}
The tier 1 file format is typically defined by the digitizer manufacturer. However, there is flexibility in choosing a different format for tier 2 files that is more conducive to pulse-shape analysis. ROOT offers not only a C++ interpreter and statistical analysis libraries but also a well-defined file format, \lstinline{.root}, optimized for efficient I/O of large datasets. This format was selected as the tier 2 file format for TOWARD and ROSA.

A quantitative study~\cite{data_formats} of various file formats, including ROOT, Protobuf, SQLite, HDF5, Parquet, and Avro, has demonstrated the superior performance of ROOT in generating statistical distributions from saved data. ROOT consistently outperformed other formats by a factor of 2 to 7 in terms of speed and achieved the smallest file size after compression.

Another important factor to consider when selecting a file format is how well it is integrated with the analysis software. The following code snippet demonstrates a typical approach to visualizing data stored in an HDF5 file using Python:
\begin{lstlisting}[language=python]
$ python
>>> import h5py
>>> import matplotlib.pyplot as plt
>>> with h5py.File('data.h5', 'r') as f:
    data = f['my_group/my_data'][:]
>>> plt.hist(data, bins=100)
>>> plt.show()
\end{lstlisting}
where \lstinline{$} represents a shell prompt, \lstinline{>>>} represents the prompt of a Python interactive session.

In contrast, visualizing data stored in the ROOT format often requires less lines of code. For example, the following commands can be used to plot the same dataset:
\begin{lstlisting}[language=c++]
$ root data.root
root [0] my_group->Draw("my_data")
\end{lstlisting}
where \lstinline{root [0]} represents the prompt of a ROOT interactive session.

ROOT's superiority lies in its seamless integration with the ROOT file format. The framework provides C++ classes and methods specifically designed to work with data stored in this format. When opening a \lstinline{.root} file, ROOT automatically recognizes datasets and provides intuitive functions for data visualization. For instance, upon opening \lstinline{data.root}, the dataset \lstinline{my_group} is automatically identified, and the \lstinline{Draw()} function creates a histogram with 100 bins to display the statistical distribution of \lstinline{my_data} on a canvas. This level of integration simplifies data analysis tasks and reduces the need for manual data manipulation.

\subsection{Flat Versus Nested Data Structure}
A common approach to storing waveform data in the ROOT format involves defining a C++ class hierarchy. For instance, one may define an \lstinline{Event} class that might contain multiple \lstinline{Waveform} objects, and each \lstinline{Waveform} could further contain several \lstinline{Pulse} objects. Setter and getter functions are defined within these classes to access and modify member variables, such as \lstinline{Waveform Event::GetWaveform(int index)} or \lstinline{double* Waveform::GetSamples()}. Objects of these classes can be saved in a ROOT file for organized data storage.

However, deeply nested data structures can introduce two challenges. First, the library containing the class definitions must be loaded by ROOT before opening the file to ensure proper access to nested data members. Second, retrieving deeply nested data can be cumbersome. For example, drawing 5 events starting from event 2 might require the following lengthy code:
\begin{lstlisting}
root [0] data->Draw("event.GetWaveform(3).GetSamples():Iteration$", "", "l", 5, 2)
\end{lstlisting}
where \lstinline{Draw()}~\cite{draw} is a member function of the ROOT \lstinline{TTree} class~\cite{tree}. It accept five parameters. The first is a string specifying the data to be drawn, using a colon ``:'' to separate the $y$ and $x$ values. \lstinline{Iteration$} is a built-in ROOT variable representing waveform sample indices. The second is a string defining data selection criteria, which is empty in this example. The third is a string specifying drawing options. ``l'' indicates connecting samples with straight lines. The fourth specifies the number of events to be drawn on the same canvas (5 in this case). The last specifies the index of the first event to be drawn.

\begin{figure}[htbp]\centering
  \includegraphics[width=0.5\linewidth]{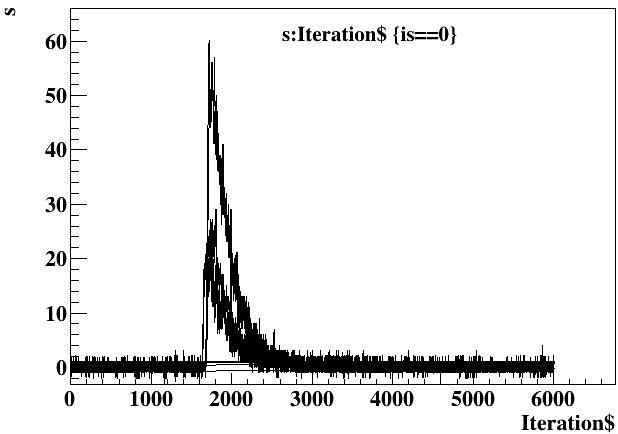}
  \caption{Waveforms drawn by \lstinline{TTree::Draw()} function. \label{f:wfs}}
\end{figure}

The same plot can be generated in TOWARD using a much shorter line of code below:
\begin{lstlisting}
root [0] t->Draw("s:Iteration$", "", "l", 5, 2)
\end{lstlisting}
The resulting plot is shown in Figure~\ref{f:wfs}.  The code is even shorter in ROSA:
\begin{lstlisting}[caption=C++ code to draw waveforms in a ROOT interactive session., label=l:st]
root [0] t->Draw("s:t", "", "l", 5, 2)
\end{lstlisting}
This reduction in code complexity is a direct result of a flat, spreadsheet-like structure, which eliminates the need for complex class hierarchies and data access methods. The use of meaningful but concise names for the spreadsheet and its columns, such as \lstinline{t} for the ROOT TTree format and \lstinline{s} for waveform samples, further enhances readability and ease of use in an interactive ROOT environment.

A potential challenge with the flat data structure arises when digitizers employ zero-suppression, a technique that discards waveform samples below a certain threshold to reduce file size. To accurately represent zero-suppressed waveforms, it is essential to record the start and end points of the suppressed regions. A \lstinline{Waveform} class can be used to store this information, with data members to track the start and end indices.

Alternatively, to maintain a simpler data structure, zero-suppressed regions can be padded with zeros. This eliminates the need to record start and end points but may increase the file size. However, ROOT's efficient compression mechanism can mitigate this impact, especially when dealing with files containing numerous repeated values. Therefore, zero-padding combined with the ROOT compression mechanism permits a simple flat data structure without significant file size penalties.

Columns of the spreadsheet saved in the TTree format can be printed with the following code:
\begin{lstlisting}[caption=Content of event 0 in ROOT TTree object \lstinline{t}.,label=l:show]
root [0] t->Show(0)
======> EVENT:0
 b   = 75.1217
 h   = 96.8783
 l   = -10.1217
 n   = 6006
 s   = 1.87833, 0.878334, 0.878334, -1.12167, -4.12167, 1.87833
 is  = 0
 db  = 2.59683
 tt  = 930
 tl  = 4683
 th  = 1207
 ttt = 3190661
\end{lstlisting}
where \lstinline{Show()} is another member function of the ROOT TTree class. It accepts an event (a row of the spreadsheet) index as its argument and displays the content of that row, including column names printed on the left.  To maintain brevity, column names are kept concise as long as they are unique.  In this example, the following column names are used:
\begin{itemize}
  \item \lstinline{b}: baseline level
  \item \lstinline{db}: baseline RMS
  \item \lstinline{h}: highest point
  \item \lstinline{l}: lowest point
  \item \lstinline{th}: time of the highest point
  \item \lstinline{tl}: time of the lowest point
  \item \lstinline{tt}: trigger time
  \item \lstinline{is}: waveform saturation indicator (Boolean)
  \item \lstinline{ttt}: trigger time tag (timestamp)
\end{itemize}
These columns represent individual numerical values. In contrast, \lstinline{s} is an array of waveform samples, with its size stored in the \lstinline{n} variable. A ROOT tree can accommodate arrays of varying sizes within different events by tracking their dimensions in a separate variable.

In ROSA, a new array, \lstinline{t}, is added in parallel with \lstinline{s}, sharing the same dimension as \lstinline{s}. It represents the time corresponding to each sample in \lstinline{s}, calculated as the array index \lstinline{Iteration$} divided by the ADC sampling frequency.

Pulse shape parameters extracted from waveform samples, such as pulse area, rise and fall time, etc., can be saved as an other flat tree in the same tier 2 file or a separate tier 3 file. The ROOT TTree class provides a mechanism to combine two spreadsheets in TTree format using its member function \lstinline{AddFriend}. The resulting combined tree can be used for analyses that require information from both tier 2 and 3 files.

\subsection{Beyond Files}
Data structures extend beyond just file formats. They encompass the broader organization of information on a computer, including file formats, internal data structures within files, directory structures, and output file naming conventions. Well-designed directory structures and clear file naming schemes can significantly simplify programming tasks. These elements should be considered as integral aspects of data structure design.

\begin{figure}[htbp]\centering
  \includegraphics[width=0.8\linewidth]{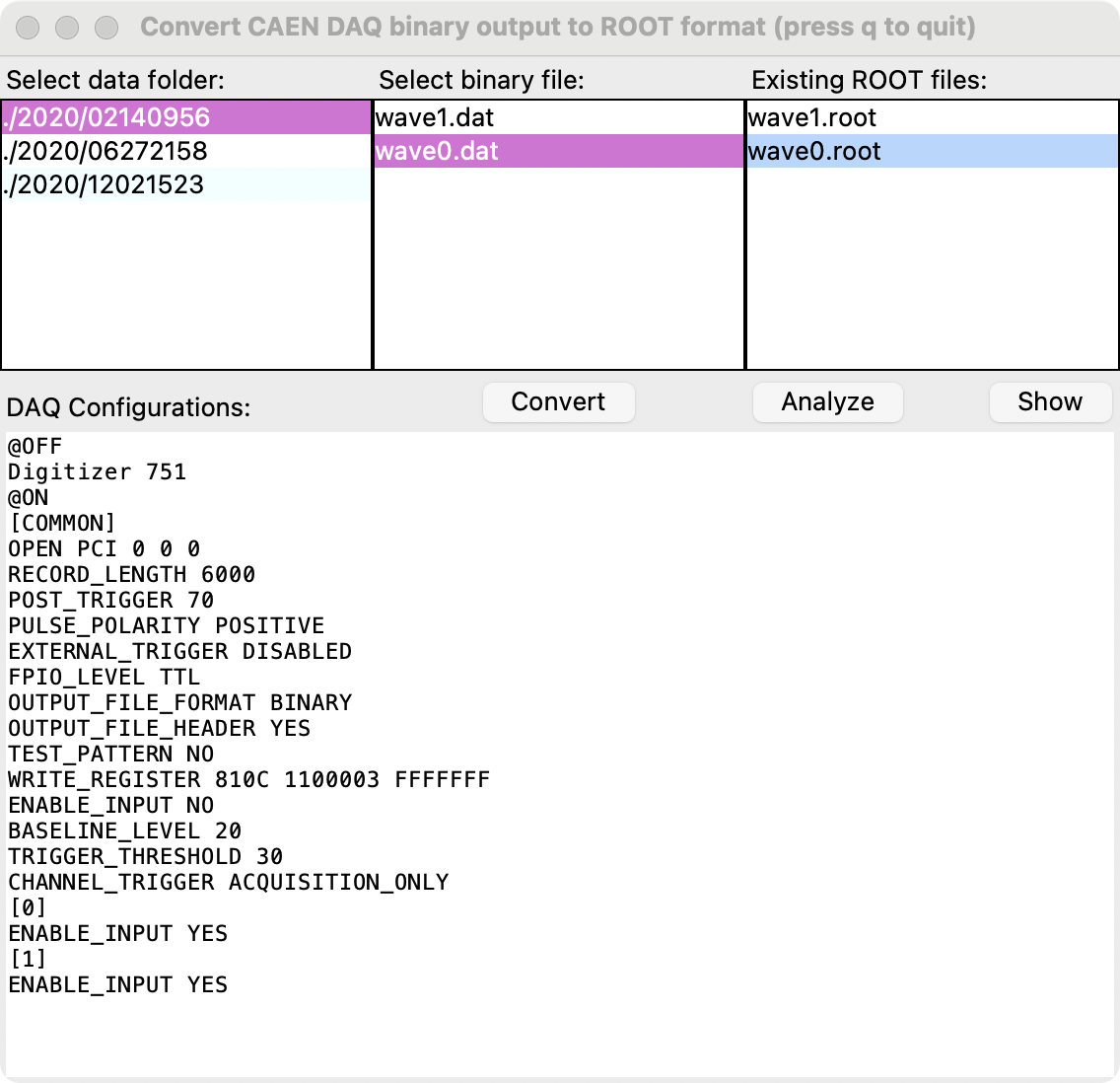}
  \caption{TOWARD's simple Python GUI, \lstinline{b2r.py}, in light mode. \label{f:b2r}}
\end{figure}

This concept is less critical for Struck digitizers, where all channel waveforms are saved within a single output file. However, software like WaveDump~\cite{wavedump}, a CAEN digitizer readout program, saves waveforms from each channel separately. For example, \lstinline{wave0.dat} represents data from channel 0, \lstinline{wave1.dat} from channel 1, and so on. These files can be further organized within folders using naming conventions such as:
\begin{itemize}
  \item \lstinline{digitizer0/} - Dedicated folder for each digitizer, 
  \item \lstinline{yyyy/mmddHHMM/} - Date and time of data acquisition, 
  \item \lstinline{yyyy/mm/dd/HHMM/} - Alternative date and time format, and
  \item \lstinline{experiment/detector1/} - Experiment specific organization.
\end{itemize}
In TOWARD, a folder containing a digitizer configuration file is automatically recognized as a data directory by its Python GUI, \lstinline{b2r.py} (Figure~\ref{f:b2r}). This configuration file can be either WaveDump's \lstinline{WaveDumpConfig.txt} or CoMPASS's \lstinline{settings.xml} (CoMPASS~\cite{compass} is another DAQ software from CAEN). The data structure encompasses the entire folder, including its name, the digitizer configuration file, and the output files from individual channels.

This overall data structure can be preserved during backup transfers from the DAQ machine to a data server using the rsync command:
\begin{lstlisting}[language=bash]
$ rsync -av data/folder server:/path/to/backup/folder
\end{lstlisting}
where the option `a' means preserve all properties of files and folders, `v' means verbose output.

\section{Graphical User Interface}
While it might seem unnecessary to develop a GUI for decoding digitizer binary files, a simple one can significantly reduce the need for command-line interactions. The following example illustrates this point.

\subsection{TOWARD GUI}
The main GUI in TOWARD shown in Figure~\ref{f:b2r} can be launched with the following command:
\begin{lstlisting}[language=bash]
$ python3 b2r.py
\end{lstlisting}
Its layout mirrors the underlying data structure. The top-left list displays sub-folders within TOWARD that contain digitizer configuration files, either \lstinline{WaveDumpConfig.txt} or \lstinline{settings.xml}. The selected folder's configuration file contents are shown in the bottom half of the GUI. Binary files generated by the digitizer within the same folder are listed in the middle. The ``Convert'' button initiates the conversion process using either the \lstinline{w2r.C} script for WaveDump binary files or the \lstinline{c2r.C} script for CoMPASS binary files. The GUI automatically collects necessary input arguments based on user selections, and the conversion progress is displayed in the terminal where the GUI is launched:
\begin{lstlisting}
  Processing w2r.c("./2020/12021523","wave0.dat",0,10,1,1000,2,12)...
  Processing event 0
  Processing event 10000
  ...
  83492 events saved in ./2020/12021523/wave0.root
\end{lstlisting}
Manually entering the command \lstinline{root w2r.c'("./2020/12021523","wave0.dat",0,10,1,1000,2,12)'} in a terminal for the same conversion task is cumbersome and prone to errors. The GUI streamlines this process, reducing complex typing to a simple button click.

\begin{figure}[htbp]\centering
  \includegraphics[width=0.9\linewidth]{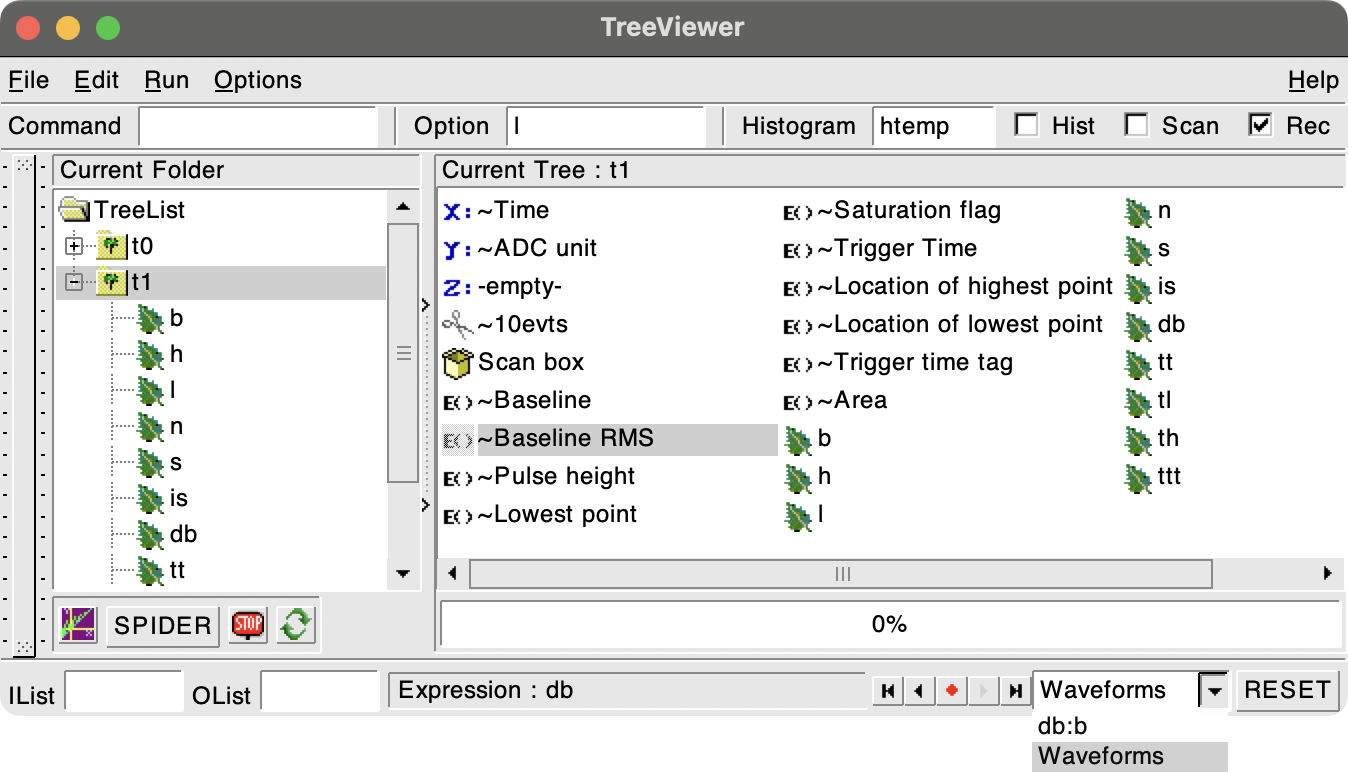}
  \caption{CERN ROOT's TreeViewer launched by clicking the ``Analyze'' button in Figure~\ref{f:b2r}. \label{f:tree}}
\end{figure}

ROOT files generated during the conversion process are displayed in the top-right list. The ``Analyze'' button initiates the \lstinline{analyze.C} ROOT script, which launches a TreeViewer as depicted in Figure~\ref{f:tree}. The \lstinline{TreeViewer} lists all trees contained within the selected data folder, renaming them from \lstinline{t} to \lstinline{t0, t1}, and so on to correspond to DAQ channel numbers.

Tree leaves are displayed both within the expanded tree structure and in the right panel alongside other items. Items with the \lstinline{E()} icon are aliases for short-named leaves. For instance, \lstinline{E()~Baseline} represents the \lstinline{b} leaf, \lstinline{E()~Baseline RMS} corresponds to the \lstinline{db} leaf, and so on. These aliases offer more descriptive labels for GUI elements while maintaining the concise leaf names suitable for command-line-based analysis.

\begin{figure}[htbp]\centering
  \includegraphics[width=\linewidth]{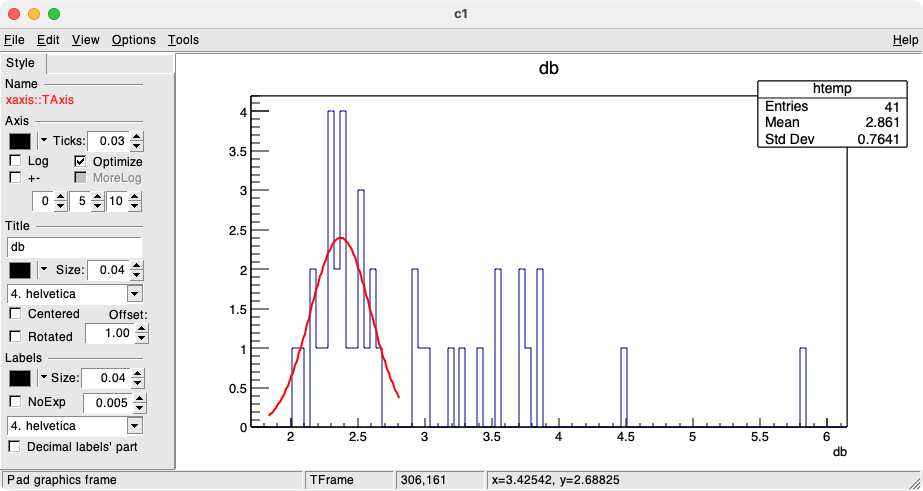}
  \caption{CERN ROOT's histogram canvas launched by double clicking the ``db'' leaf in Figure~\ref{f:tree}. \label{f:db}}
\end{figure}

Double-clicking on a leaf or its alias in the \lstinline{TreeViewer} launches an interactive histogram as shown in Figure~\ref{f:db}. Users can customize the appearance of the histogram, perform fitting operations, and query statistic properties through mouse clicks and drags.

Complex plotting processes can be pre-defined or recorded as macros and listed in the bottom-right pull-down menu of the TreeViewer. Figure~\ref{f:tree} shows two predefined macros: \lstinline{db:b} and Waveforms. The \lstinline{db:b} macro generates a 2D histogram as a scatter plot with the \lstinline{y}-axis representing baseline RMS (\lstinline{db}) and the \lstinline{x}-axis representing baseline value (\lstinline{b}). The Waveforms macro overlays waveforms from a few initial events on a canvas for rapid data quality assessment.

\begin{figure}[htbp]\centering
  \includegraphics[width=0.8\linewidth]{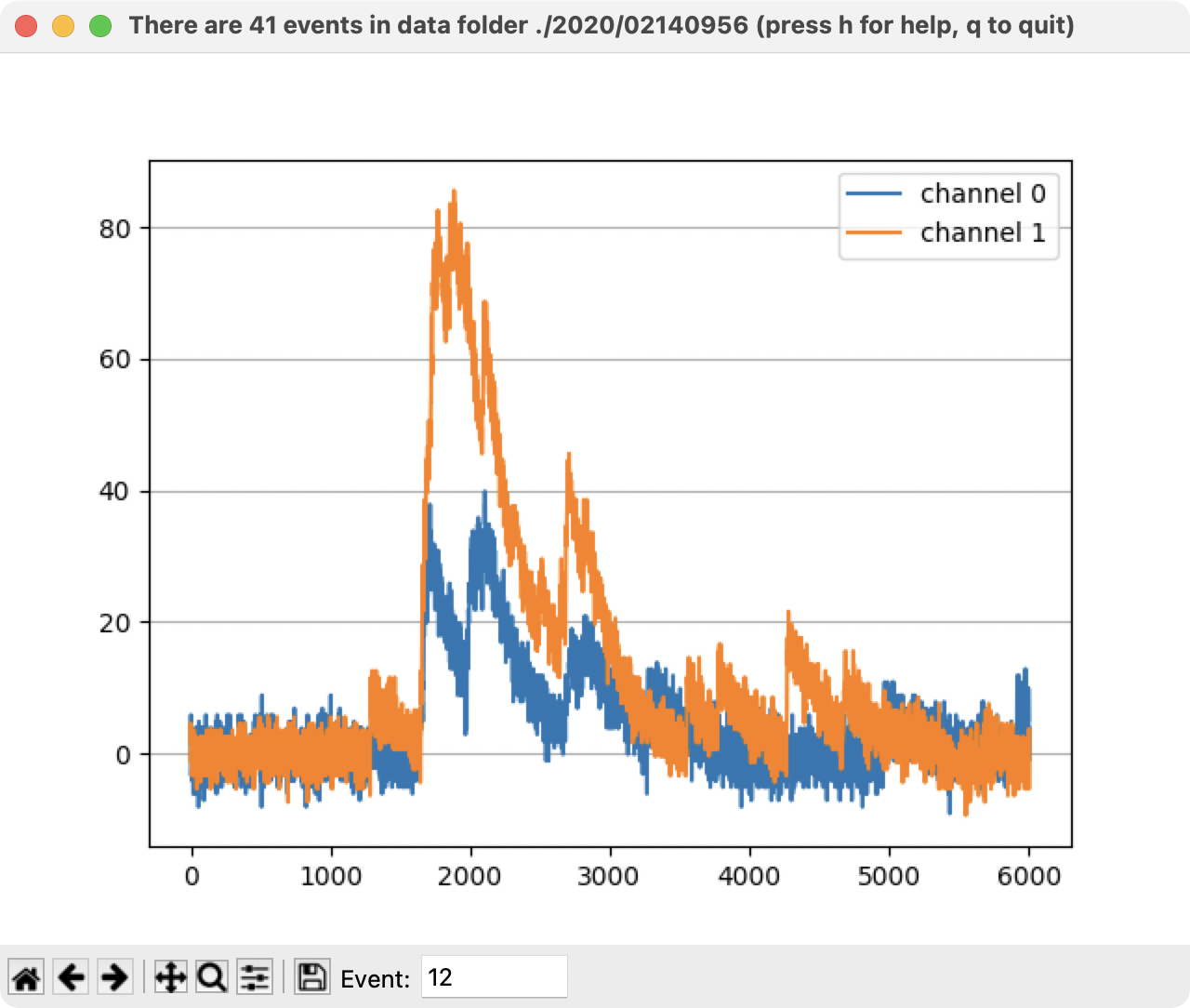}
  \caption{TOWARD's simple matplotlib canvas created by \lstinline{show.py} for sequential event display. \label{f:wf}}
\end{figure}

The \lstinline{TreeViewer} is specifically designed for visualizing statistic distributions of a large number of events. A Python script \lstinline{show.py} is included in TOWARD, which allows users to examine individual events sequentially using a matplotlib canvas (Figure~\ref{f:wf}). This canvas can be launched by clicking the ``Show'' button under the ROOT file list in TOWARD's main GUI shown in Figure~\ref{f:b2r}. Pressing the \lstinline{h} key when the canvas window is active will print the following help message in the terminal where the main GUI is launched:
\begin{lstlisting}
  --- List of key bindings ---
0-7:        toggle channel 0 to 7
<Space>:    next event
b:          previous event
q:          quit
\end{lstlisting}
The canvas provides a quick and efficient way to visually inspect individual events, similar to the output generated by the code in Listing~\ref{l:st}. This interactive visualization is ideal for rapid data quality checks. For more in-depth analysis of individual events, the C++ code demonstrated in Listing~\ref{l:st} can be utilized.

\subsection{ROSA GUI}
\begin{figure}[htbp]\centering
  \includegraphics[width=\linewidth]{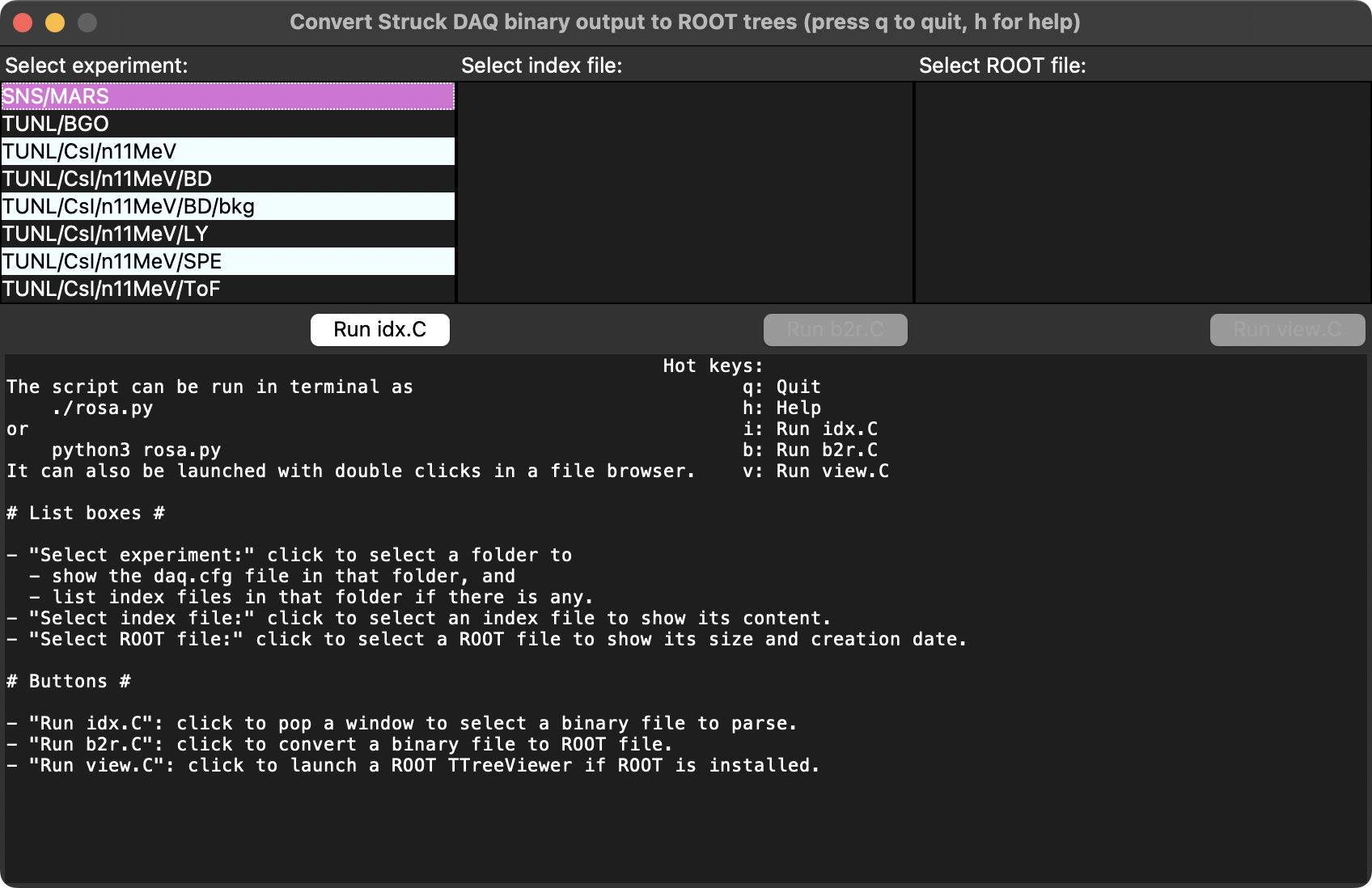}
  \caption{ROSA's simple Python GUI, \lstinline{rosa.py}, in dark mode. \label{f:rosa}}
\end{figure}

ROSA also offers a user-friendly GUI, accessible by running \lstinline{python3 rosa.py}. Figure~\ref{f:rosa} showcases ROSA's dark mode theme. Both TOWARD and ROSA GUI adapt their light/dark themes to match the user's operating system settings.

The bottom half of the ROSA GUI initially displays a help message. Upon selecting a data folder from the top-left list, the content of the DAQ configuration file (\lstinline{daq.cfg}) within that folder replaces the help message. A typical \lstinline{daq.cfg} file is shown in Listing~\ref{l:daq} and will be further explained in Section~\ref{s:daqcfg}.

The \lstinline{Run idx.C} button launches a file selection window. Users can then select a Struck DAQ binary output file, which will be passed as input to the ROOT script \lstinline{idx.C}. This script reads the header of the binary file, verifies its consistency with \lstinline{daq.cfg}, and writes the starting positions and lengths of data blocks within the binary file into a CSV file stored in the selected data folder. Upon selecting a generated CSV file in the middle list, the contents of that file will replace the contents originally in the bottom half of the GUI.

A key distinction between TOWARD and ROSA lies in their handling of binary output files. TOWARD assumes that these files are located in the same folder as the DAQ configuration file, mirroring WaveDump's behavior. In contrast, ROSA allows for binary files to be stored in separate folders.

ROSA associates CSV files generated by \lstinline{idx.C} with the DAQ configuration file. These CSV files are listed in the top-middle section of the ROSA GUI. The \lstinline{Run b2r.C} button is enabled only when a CSV file is present in the folder. Clicking this button triggers the conversion of the corresponding binary file to a ROOT file, utilizing the information stored in the CSV file. The generated ROOT files are saved within the same folder and displayed in the top-right list.

If ROOT files are detected, the \lstinline{Run view.C} button becomes active. Clicking this button launches a TreeViewer, similar to the one depicted in Figure~\ref{f:tree}.

\section{DAQ Configuration Files}
\label{s:daqcfg}
Digitizer configurations used to generate binary files serve as metadata for waveform data.  CAEN DAQ software, including WaveDump and CoMPASS~\cite{compass}, stores configuration files in \lstinline{WaveDumpConfig.txt} and \lstinline{settings.xml}, respectively. Both are human-readable text files.

When multiple configuration files exist in different locations, settings from the file in the DAQ software's launch folder take precedence. TOWARD hence recommends the following steps to ensure proper organization of metadata and data:
\begin{enumerate}
  \item Create a new experiment folder structure.
  \item Copy an example DAQ configuration file into the experiment folder.
  \item Modify the configuration file to match experiment requirements.
  \item Launch the DAQ software within the same folder to acquire data.
\end{enumerate}
By adhering to this process, tier 1 data and DAQ configurations are naturally organized within the desired folder structure, aligning with the principles outlined in Section~\ref{s:struct}.

Unlike CAEN DAQ software, Struck DAQ software stores its configuration within a ROOT file as a C++ class object. This dependency on the DAQ software can limit the flexibility of binary decoding code if configuration files are stored alongside binary files. To maintain independence, ROSA introduces a simplified DAQ configuration file, \lstinline{daq.cfg}, written in plain English for easy understanding (see Listing~\ref{l:daq}).

\begin{lstlisting}[caption=Content of customized DAQ configuration file daq.cfg in ROSA/SNS/MARS/ folder., label=l:daq]
experiment: SNS/MARS
daq: SIS3316
bit: 14
sampling rate: 250 MHz
number of channels per module (card/board): 16
number of modules used: 2
local channel id:
   0    1    2    3    4    5    6    7    8    9   10   11   12   13   14   15
   0    1    2    3    4    5    6    7    8    9   10   11   12   13   14   15
global channel id (card id x number of channels per card + local channel id):
   0    1    2    3    4    5    6    7    8    9   10   11   12   13   14   15
  16   17   18   19   20   21   22   23   24   25   26   27   28   29   30   31
channel status (1: connected, 0: empty):
   1    1    1    1    1    1    1    1    1    1    1    1    1    1    1    1
   1    0    1    0    0    0    0    0    0    0    0    0    0    0    0    0
channel sync requirement (1: sync, 0: no need to sync):
   1    1    1    1    1    1    1    1    1    1    1    1    1    1    1    1
   0    0    0    0    0    0    0    0    0    0    0    0    0    0    0    0
\end{lstlisting}

Due to a firmware bug in the Struck digitizers used in the author's experiments, some channels recorded additional waveforms. The final three lines of \lstinline{daq.cfg} are used to instruct the binary decoding code to perform channel alignment, compensating for this inconsistency.

The introduction of a customized DAQ configuration file eliminates the need to retain tier 1 data (binary files) generated by Struck digitizers alongside ROSA. Tier 2 data, consisting of ROOT files converted from binary files, are maintained within the same folder as \lstinline{daq.cfg}.

\section{Scripts}
As highlighted in Section~\ref{s:hvl}, C++ code in TOWARD and ROSA is structured as ROOT scripts. Similar to Python and Bash scripts, these C++ scripts can be executed without manual compilation, provided CERN ROOT is installed on the system. Each script focuses a few simple tasks, allowing for chaining to perform more complex operations.  This approach contrasts with traditional C++ programming practices, where small tasks related to PSA will be implemented as member functions within classes like \lstinline{Waveform} or \lstinline{Pulse}.

The decision to use scripts instead of classes was intentional. Common PSA algorithms are widely available in various languages, including C, C++, Python, Julia, and others. Reimplementing these algorithms as member functions within C++ classes would be redundant. TOWARD and ROSA concentrate on providing scripts for fundamental pulse processing tasks, reinforcing simple, flat ROOT TTree file format. The resulting ROOT files can be accessed without loading any custom C++ libraries. Notably, such simple ROOT trees can be easily read using Python's \lstinline{uproot}~\cite{uproot} package or Julia's \lstinline{UnROOT}~\cite{unroot} package, eliminating the dependency on CERN ROOT libraries.

\subsection{Binary File Decoding Scripts}
The data structure of binary output files from digitizers can vary between manufacturers. Consequently, a separate decoding script is typically required for each type of binary file. For example,
\begin{itemize}
  \item \lstinline{TOWARD/w2r.C} is a ROOT script to decode binary files from CAEN DAQ software, WaveDump,
  \item \lstinline{TOWARD/c2r.C} is a ROOT script to decode binary files from CAEN DAQ software, CoMPASS.
  \item \lstinline{ROSA/idx.C} is a ROOT script to index binary files from a private Struck DAQ software, ngmdaq,
  \item \lstinline{ROSA/b2r.C} is a ROOT script to decode binary files from ngmdaq.
\end{itemize}

Beyond decoding binary files and saving waveforms in ROOT TTree format, the \lstinline{TOWARD/w2r.C} and \lstinline{TOWARD/c2r.C} scripts perform additional tasks. These include calculating the baseline of each waveform, adjusting it to zero, and recording the highest and lowest points within the waveform. Additionally, they identify the location where the first pulse exceeds a predefined threshold.

\subsection{PSA Scripts}
While TOWARD and ROSA are not designed to encompass all sophisticated PSA algorithms, they do include some for common tasks and experiment-specific purposes. For example,
\begin{itemize}
  \item \lstinline{TOWARD/q2i.C} is a ROOT script to get heights of charge pulses using a trapezoidal filter and convert charge pulses to current ones using numerical differentiation,
  \item \lstinline{TOWARD/i2q.C} is a ROOT script to convert current pulses to charge ones,
  \item \lstinline{TOWARD/integrate.C} is a ROOT script to integrate waveforms in a certain range and save the result to the original ROOT tree,
  \item \lstinline{ROSA/TUNL} folder contains many ROOT scripts for various purposes, such as calculating PSD parameters for liquid scintillation detectors, correcting overshooting after a large pulse, calculating time difference between pulses from different channels, etc.
\end{itemize}

\subsection{Batch Scripts}
ROSA provides Bash scripts, like \lstinline{b2r.sh} and \lstinline{idx.sh}, for submitting batch jobs to High-Performance Computing clusters. Taking \lstinline{b2r.sh} as an example, this script iterates through CSV files within a designated data folder. For each CSV file, it submits a job for decoding the associated binary file. The script then monitors and displays the progress of these jobs until completion. These scripts are ideal for processing large datasets efficiently, while the GUI excels at handling a smaller number of files interactively.

\section{Package Maintenance and Distribution}
Both TOWARD and ROSA are publicly available as free, open-source packages under the MIT license on GitHub. Each package utilizes README.md files at the top level and within subdirectories to document various aspects. These Markdown-formatted files can be opened directly in any text editor and are automatically rendered into well-formatted HTML pages on GitHub for easy online browsing. These files reside alongside the scripts within the packages, accessible both online and offline.

These packages are remarkably small in size, primarily consisting of plain-text scripts. This makes them ideal for co-location with the data they process, promoting reproducibility of analysis results. By leveraging Git's version control system, the exact code version used for a specific analysis can be precisely retrieved based on the timestamps of the resulting data files.

Both the data structure and analysis code are deliberately designed to be lightweight and version-agnostic. This minimizes the need to maintain various versions of data and analysis code throughout an experiment's lifecycle, significantly reducing overhead.

\section{Conclusion}
Digital pulse processing is a valuable tool for extracting information from various signals. This paper highlights practical approaches to digital pulse processing, focusing on simplicity and efficiency.

We advocate for a balanced approach to software design, avoiding both excessive modularization and overly generic frameworks. Overly modularized systems can introduce complexity and inefficiencies, while overly generic frameworks may be difficult to maintain. Striking the right balance is essential for creating effective and maintainable software.

We emphasize the use of flat, simple data structures instead of complex nested C++ classes. This simplifies data handling and reduces the need for additional libraries.

By leveraging the ROOT C++ interpreter, cling, we avoid the overhead of compiling C++ scripts while maintaining reasonable performance. The ROOT interpreter offers a convenient environment for interactive data analysis and exploration.

The ROOT TTree::Draw function works seamlessly with the short variable names used in our flat tree structure. This streamlined approach simplifies data visualization and analysis tasks, making it easier to extract meaningful insights from the data.

We also advocate for a balanced approach to user interfaces, combining command-line tools with graphical interfaces. This provides flexibility and caters to different user cases and needs.

By adopting these practical strategies, researchers can confidently apply digital pulse processing techniques to their specific domains without being overly concerned about the theoretical correctness of their code.

\section{Acknowledgements}
The work is partially supported by National Science Foundation (NSF) award PHY-2411825 and Department of Energy award HEP DE-SC0022167.

I would like to thank Gemini, a large language model developed by Google AI, for its invaluable contributions in refining the language and improving the clarity of this paper.

\bibliography{ref}{}
\bibliographystyle{unsrt} 

\end{document}